\documentclass[twocolumn,showpacs,preprintnumbers,amsmath,amssymb]{revtex4-1}

\usepackage{graphicx}
\usepackage{dcolumn}
\usepackage{bm}

\usepackage{epsfig}

\begin{document}

\title{Comment on ''Counterintuitive consequence of heating in strongly driven intrinsic junctions of Bi$_2$Sr$_2$CaCu$_2$O$_{8+\delta}$ mesas.}

\author{V. M. Krasnov}
\affiliation{
Department of Physics, Stockholm University, AlbaNova University Center, SE-10691 Stockholm, Sweden
}

\date{\today}

\begin{abstract}
C. Kurter et. al., \cite{Kurter} analyzed acute self-heating in
large Bi$_2$Sr$_2$CaCu$_2$O$_{8+\delta}$ (Bi-2212) mesa
structures. They attributed observed peaks in conductance to
heating of mesas up to the superconducting critical temperature
$T_c$, extrapolated this statement to much smaller mesas, used in
Intrinsic Tunnelling Spectroscopy (ITS), and called for
reinterpretation of ITS data. They also suggested a universal
``figure of merit" for the shape of tunnelling characteristics.
Here I argue that the peak in c-axis conductance usually occurs
well below $T_c$; that for small ITS mesas with small self-heating
the peak represents the superconducting gap; and that the genuine
shape of tunnelling characteristics for cuprates is not universal
but depends on doping, uniformity and geometry.

\end{abstract}

\pacs{
74.72.Gh, 
74.55.+v, 
74.50.+r, 
74.72.Kf. 
}

\maketitle

Self-heating in ITS is actively discussed \cite{HeatCom} for more
than a decade (see Ref. \cite{SecondOrder} for a recent summary).
ITS appeared as a result of obviation of self-heating either via
short pulse measurements [8a]$^{[1]}$ or miniaturization
[10a,15b,9,7]$^{[1]}$, \cite{Winkler,KrPhysC,Latysh,SecondOrder}
(in what follows the subscript $^{[1]}$ indicates citations from
the reference list of \cite{Kurter}). Clearly, it is hard to
decipher experimental data if junctions are heated above $T_c$
below the sum-gap voltage $V_g=2\Delta/e$, where $\Delta$ is the
superconducting energy gap. But if $T(V_g)$ is significantly
smaller than $T_c$, the spectroscopic information can be obtained.

Self-heating in mesa structures was thoroughly studied (primarily
in microelectronics) and is well understood. It depends on
materials parameters and mesa geometry and decreases in smaller
mesas [15b,9]$^{[1]}$. Therefore, size-dependence, or independence
provides an unambiguous clue about heating, or spectroscopic
origin of observed features [9]$^{[1]}$. This is the subject of
paper \cite{Kurter}. But unlike previous ITS studies, which were
focusing on small micron-to-submicron mesas
[9,7]$^{[1]}$,\cite{SecondOrder} Kurter et al consider large
300-10 $\mu$m mesas. They confirmed that self-heating is reduced
with decreasing mesa sizes.

The main quantitative statement of Ref. \cite{Kurter} is that
sharp peaks in $dI/dV$ represent a transition to the normal state,
$T_m=T_c$. However, this is in conflict with their own
estimations: applying their thermal resistance $\alpha=70$K/mW for
the $I-V$ curve at $T_B=4.8K$ from Fig.7, we obtain the effective
temperature at the peak ($I=0.7$mA, $V=19*33$ mV) $T_m =T_B+
\alpha IV \simeq 35.5$K, which is less than a half of $T_c=74K$.
There are also many direct experimental evidences against this
statement:

i) In large mesas, the (infinitely) sharp peak corresponds to an
onset of back-bending of $I-V$ curves as a result of self-heating.
However, multiple quasiparticle (QP) branches, due to one-by-one
switching of intrinsic Josephson junctions in the mesa from the
superconducting to the QP state, can still be observed in the
back-bending region (see Fig. 1 in Ref. \cite{HeatCom} and Fig. S2
in the supplementary to [18]$^{[1]}$). Consequently, the mesas are
remaining well below $T_c$ not only at the peak but also at
significantly larger bias, corresponding to dissipation powers
several times that at the peak.

ii) Presence of the ac intrinsic Josephson effect in the
superconducting state leads to emission of electromagnetic waves
from Bi-2212 mesas. For similar large Bi-2212 mesas as in
Ref.\cite{Kurter}, significant emission was observed (see Fig. 2
in [18]$^{[1]}$) not only at the onset of back-bending ($IV \simeq
10$ mW) but also in the whole back-bending region up to $25$ mW.

iii) Direct measurement of mesa temperatures for small ITS mesas
showed that at the peak $T_m$ remains well below $T_c$
[9]$^{[1]}$. Recently this conclusion was confirmed by
''intrinsic" detection of $T_m$ \cite{HeatCom,SecondOrder}. In
this case, $T-$dependent characteristics of the mesa itself were
used for determination of $T_m$, avoiding any possible lag between
the mesa and the thermometer.

For small mesas, used in ITS, there are many evidences that the
peak not only occurs well below $T_c$ but also represents the
superconducting gap $\Delta$:

iv) A clear evidence comes from observation of large
magnetoresistance (which is the signature of superconductivity and
disappears at $T>T_c$) not only at the peak but also at bias well
above the peak [33]$^{[1]}$,\cite{Lee}. The way in which the peak
is suppressed by magnetic field allows an unambiguous association
of the peak in small ITS mesas with the sum-gap tunnelling
singularity \cite{PhC2004}.

v) Additional dips are observed at twice the peak voltage
\cite{SecondOrder}. They are due to non-equilibrium effects and
confirm the gap origin of the peak [10b]$^{[1]}$,\cite{Cascade}.

vi) The way in which the peak is affected by self-heating is
consistent with the gap origin of the peak \cite{SecondOrder}.

vii) Size-independence of ITS characteristics of small mesas was
proven [9,7]$^{[1]}$. This unambiguously indicates that the peak
voltage per intrinsic junction is a material property, $\Delta$,
rather than an artifact of self-heating.

viii) In break junctions strong phonon resonances were observed up
to the sharp peak \cite{Ponomarev}. Since their amplitude is
proportional to the Josephson current squared, this not only
precludes heating above $T_c$ at the peak, but also allows
unambiguous attribution of the sharp peak to the sum-gap
singularity with the corresponding Riedel singularity of the
supercurrent amplitude \cite{Ponomarev}.

Kurter et al. also suggested a universal ``figure of merit" for
tunnelling characteristics of cuprates: ``correct" spectra should
look like their mechanical contact (MCT) characteristics, i.e.,
have a non-sharp peak followed by a dip/hump. In support they
claim that ``there is excellent agreement" among MCT, scanning
tunnelling (STS) and angle-resolved photoemission (ARPES)
spectroscopies ``with no significant discrepancies". However,
there exist significant differences among those three surface
tunnelling techniques: for example, STS shows (almost) no
$\Delta(T)$ dependence at $T_c$ [27]$^{[1]}$, MCT shows BCS-like
$\Delta(T)$ [20,21]$^{[1]}$, while ARPES shows BCS-like
$T-$dependence only in the nodal Fermi arc region \cite{Shen}.
Also the doping dependence of the pseudogap (PG) is different:
according to STS the PG persists [27]$^{[1]}$, while according to
MCT it ceases to exist [21]$^{[1]}$ in overdoped Bi-2212. ARPES
sees a gradual expansion of the PG-free Fermi arc region with
overdoping \cite{Shen}. For comparison, ITS shows BCS-like
$T-$dependence of the bulk gap \cite{SecondOrder} and indicates
vanishing of the PG at the critical point on the overdoped side of
the doping diagram \cite{Doping}.

It is well known that tunnelling in cuprates crucially depends on
QP momentum due to the $d-$wave symmetry of $\Delta$ \cite{dwave},
strong momentum dependence of the tunnelling matrix \cite{Millis},
and on coherence (momentum conservation) of tunnelling
[8b]$^{[1]}$. There are substantial differences in tunnelling
geometry for the considered spectroscopic techniques: for STS this
is [CuO/BiO-vacuum-normal metal] surface $c-$axis tunnelling; for
MCT this is [CuO/BiO-vacuum-BiO/CuO] surface tunnelling in an
unknown direction; and for ITS this is [CuO-BiO-CuO] $c-$axis
tunnelling in a bulk single crystal. Variation of electronic
structure and doping level at the surface [CuO/BiO-vacuum] may
cause differences between surface and bulk characteristics
\cite{SecondOrder}. Therefore, any universal shape for all
tunnelling characteristics, irrespective of experimental details,
should not be expected.

The lack of universality leads to a variety of characteristics
reported in point contacts and break-junctions on cuprates. In
particular, Ponomarev et al., \cite{Ponomarev} have observed
characteristics remarkably similar to ITS with sharp $dI/dV$ peaks
and strong phonon resonances. 
The latter implies that the crystal structure was intact and
tunnelling was indeed of the intrinsic-type. Theoretically, very
sharp sum-gap peaks are expected in case of coherent QP tunnelling
[8b]$^{[1]}$. This is supported by observation of large $I_c R_n
\sim \Delta/e$ both in ITS \cite{Doping} and in MCT junctions
[24a]$^{[1]}$, which for $d-$wave superconductors is only possible
in case of coherent tunnelling \cite{dwave}.

Furthermore, it is known that the sharpness of QP peaks is
crucially dependent on doping. From ITS and ARPES data it follows
that the peak is intrinsically sharp in overdoped Bi-2212 and
rapidly loosing sharpness, becoming very similar to MCT
characteristics, with underdoping \cite{Doping}. However,
observation of the MCT-like spectra does not imply that they are
free from artifacts. In fact, the ``correct" behavior reported in
Ref. [3]$^{[1]}$ cited by Kurter et. al., is most likely an
artifact of switching of underlaying junctions below the mesa in
the resistive state. In fact not only one, but multiple dip/humps
could be observed \cite{Winkler,YouAPL}. The corresponding small
sub-branches can be seen in Fig. 3 of Ref. [3]$^{[1]}$ at $V\sim$
800mV. This artifact leads to strong underestimation of $\Delta$
\cite{Kurter}. It disappears in smaller mesas due to decrease of
area to perimeter ratio \cite{SecondOrder}. There are also other
artifacts that can strongly reduce the sharpness of the measured
peak, such as non-uniformity of junctions in the mesa
\cite{KrPhysC,SecondOrder}, bad contact resistance at the top of
the mesa and variation of doping within the mesa. The latter is
specifically important for sub-micron mesas for which the
influence of the insulating (undoped) edges of the mesa is growing
proportionally to the perimeter-to-area ratio.

To conclude, it is clear that self-heating distorts $I-V$
characteristics and in case of large mesas may completely preclude
spectroscopic studies. Yet, association of the infinitely sharp
peak in $dI/dV$ with heating up to $T_c$ is incorrect.
Self-heating is decreasing with the mesa size. For small enough
mesas, used in recent ITS studies, the peak in $dI/dV$ represents
the superconducting gap. This conclusion was crosschecked in many
ways, as described above, and does not need reinterpretation. The
genuine shape of $dI/dV$ characteristics remains an unsettled
issue because the sharpness of the sum-gap peak can indeed be
enhanced by self-heating, or reduced by inhomogeneity of junctions
and doping distribution in the mesa. In any case, it does strongly
depend on doping and the tunnelling geometry. Therefore, any
universal shape for all types of cuprate tunnel junctions is not
to be expected.

\end {document}